\documentclass[12pt]{article}

\usepackage{scicite}

\usepackage{times}
\usepackage{graphicx}

\newcommand{\bra}[1]{\langle #1|}
\newcommand{\ket}[1]{|#1\rangle}

\def\ex{{\mathbf e}_x}                            
\def\ey{{\mathbf e}_y}                            
\def\ez{{\mathbf e}_z}                            
\def\es{{\mathbf e}_s}                            
\def\nm{{\ {\rm nm}}}						
\def\Rb87{^{87}\rm{Rb}}					
\def\Hz{{\ {\rm Hz}}}						
\def\MHz{{\ {\rm MHz}}}						
\def\ms{{\ {\rm ms}}}						
\def\us{{\ \mu{\rm s}}}						

\newenvironment{sciabstract}{%
\begin{quote} \bf}
{\end{quote}}

\newcounter{lastnote}
\newenvironment{scilastnote}{%
\setcounter{lastnote}{\value{enumiv}}%
\addtocounter{lastnote}{+1}%
\begin{list}%
{\arabic{lastnote}.}
{\setlength{\leftmargin}{.22in}}
{\setlength{\labelsep}{.5em}}}
{\end{list}}

\title{Visualizing edge states with an atomic Bose gas in the quantum Hall regime}

\author
{B. K. Stuhl$^{1,\ast}$, H.-I Lu$^{1,\ast}$, L. M. Aycock$^{1,2}$, D. Genkina$^1$, and I. B. Spielman$^{1,^\dagger}$\\
\\
\normalsize{$^{1}$Joint Quantum Institute}\\
\normalsize{National Institute of Standards and Technology, and University of Maryland}\\
\normalsize{Gaithersburg, Maryland, 20899, USA}\\
\normalsize{$^{2}$Cornell University}\\
\normalsize{Ithaca, New York, 14850, USA}\\
\\
\normalsize{$^\ast$These authors contributed equally to this work.}\\
\normalsize{$^\dagger$To whom correspondence should be addressed; E-mail:  ian.spielman@nist.gov.}
}

\date{}

\begin{document} 


\baselineskip24pt


\maketitle 


\begin{sciabstract}
We engineered a two-dimensional magnetic lattice in an elongated strip geometry, with effective per-plaquette flux $\approx4/3$ times the flux quanta.  We imaged the localized edge and bulk states of atomic Bose-Einstein condensates in this strip, with single lattice-site resolution along the narrow direction.  Further, we observed both the skipping orbits of excited atoms traveling down our system's edges, analogues to edge magnetoplasmons in 2-D electron systems\cite{Kern1991,Ashoori1992}, and a dynamical Hall effect for bulk excitations\cite{LeBlanc2012}.  Our lattice's long direction consisted of the sites of an optical lattice and its narrow direction consisted of the internal atomic spin states\cite{Celi2014,Mancini2015}.  Our technique has minimal heating, a feature that will be important for spectroscopic measurements of the Hofstadter butterfly\cite{Hofstadter1976,Celi2014} and realizations of Laughlin's charge pump\cite{Laughlin1981}.
\end{sciabstract}

\paragraph*{Introduction}

In materials the quantum Hall effects represent an extreme quantum limit, where a system's behavior defies any description with classical physics.  In the modern parlance, the integer quantum Hall effect (IQHE) for two-dimensional (2-D) electronic systems in magnetic fields\cite{Klitzing1980} was the first topological insulator\cite{Hasan2010}: a bulk insulator with dispersing edge states --- always present in finite-sized topological systems--- which give rise to the IQHE's signature quantized Hall resistance\cite{Thouless1982}.  


In classical systems the magnetic field acts purely through the Lorentz force, while in quantum systems a particle with charge $q$ in a uniform field $B$ additionally acquires an Aharonov-Bohm phase $\phi_{\rm AB}/2\pi = \mathcal{A} B / \Phi_0$ after its path encircles an area $\mathcal{A}$ normal to $B$.  (Here $\Phi_0=2\pi\hbar/q$ is the flux quanta and $2\pi\hbar$ is Planck's constant.)  We engineered a 2-D lattice with square geometry for ultracold charge-neutral atoms in which we directly controlled the acquired phases as atoms traversed the lattice, giving a tunneling phase $\phi_{\rm AB}/2\pi \approx4/3$ around each plaquette.  These phases take the place of the Aharonov-Bohm phases produced by true magnetic fields and suffice to fully define the effective magnetic field.  Aharonov-Bohm phases of order unity ---only possible in engineered materials\cite{Geisler2004,Hunt2013}, or in atomic\cite{Jaksch2003,Aidelsburger2013,Miyake2013,Jotzu2014,Aidelsburger2014,Mancini2015} and optical\cite{Hafezi2013} settings--- fragment the low-field Landau levels into the fractal energy bands of the Hofstadter butterfly\cite{Hofstadter1976}.   Such Hofstadter bands are generally associated with a non-zero topological index: the Chern number\cite{Thouless1982}.



Topologically nontrivial bulk properties are reflected by the presence of edge channels, composed of edge states, with quantized conductance.  In fermionic systems, the number of edge channels is fixed by the aggregate topological index of the filled bands\cite{Thouless1982,Hasan2010,Beugeling2012}; this ultimately gives rise to phenomena such as the IQHE for electrons.    Conceptually the constituent edge states can be viewed as skipping orbits\cite{Buttiker1988PRB,Hasan2010,Montambaux2011}: in the presence of a strong magnetic field, nascent cyclotron orbits near the boundary reflect from the hard wall before completing an orbit, leading to skipping trajectories following the system's boundary.  In contrast, localized bulk states correspond to closed cyclotron orbits.


By applying large effective fields to atomic Bose-Einstein condensates (BECs), we directly imaged individual, deterministically prepared, bulk and edge eigenstates.  In IQHE systems these states would govern the conductivity, but as individual eigenstates they exhibit no time-dependance.  The corresponding dynamical entities are edge magnetoplasmons, consisting of superpositions of edge eigenstates in different Landau levels\cite{Kern1991,Ashoori1992}, or here magnetic bands.  We launched these excitations and recorded their full motion for the first time, observing both a chiral drift along the system's edge and the underlying skipping motion.

\begin{figure}[h!]
  \centering
    \includegraphics[width=3.4in]{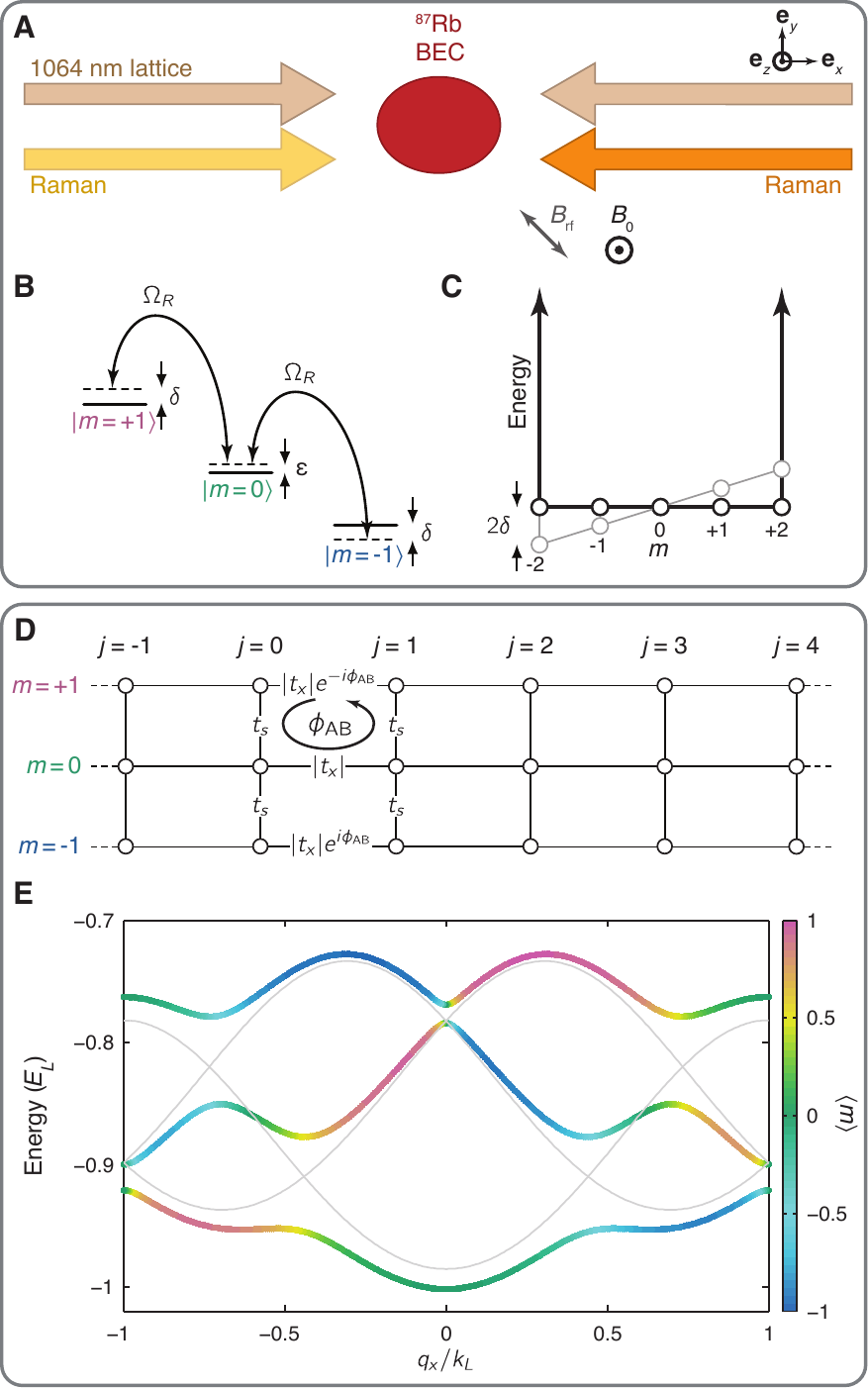}
    \caption{Hybrid 2-D lattice. (\textbf{A}) $N\approx 10^5$ atom $^{87}$Rb BECs, confined in a 1-D optical lattice with a bias magnetic field $B_0\ez$, were illuminated either by a pair of counter-propagating Raman lasers or an rf magnetic field $B_{\rm rf}$.  (\textbf{B}) The Raman or rf field linking the three internal $\ket{m}$ states with Rabi frequency $\Omega_R$ was detuned from the $g \mu_B B_0/h \approx 0.817\MHz$ or $1.35\MHz$ Zeeman splitting by $\delta$.  The corresponding quadratic Zeeman shift additionally lowered $\ket{m=0}$ by $\epsilon = 0.05 E_L$ or $0.13E_L$.  The Raman lasers' relative phase was actively stabilized at a beam-combiner adjacent to the optical lattice retroreflection mirror.  (\textbf{C}) The lattice along $\es$ can be considered as a square well with hard walls at $m = \pm2$ for which $\delta \neq 0$ tilts the potential towards one wall or the other. (\textbf{D}) The 2-D hybrid lattice, where the non-spatial dimension is built from the internal states $\ket{m}$ with an effective magnetic flux per plaquette $\Phi/\Phi_0=\phi_{\rm AB}/2\pi$. (\textbf{E})  Three lowest magnetic bands (rainbow colors), computed for our full lattice without making the tight binding approximation, with parameters ($\hbar\Omega_R,\,V,\,\delta,\,\epsilon)=(0.2,6,0,0.05)E_L$.  The pale curves were computed for $\hbar\Omega_R = 0$.}
    \label{Setup}
\end{figure}

The Harper-Hofstadter Hamiltonian
$$
H = -\sum_{j,m} \Big[t_{x}\ket{j+1,m}\bra{j,m} + t_{s}\ket{j,m+1}\bra{j,m} + \mathrm{h.c.} \Big],\ \ \ \ \ \ \ \ \ \ \ \ (1)
$$
with complex hopping amplitudes $t_x$ and $t_s$ governs the motion of charged particles moving in a 2-D lattice\cite{Harper1955,Hofstadter1976} with sites labeled by $j$ and $m$, the situation which we engineered for our neutral atoms. Analogous to the Landau gauge in continuum systems, we describe our experiment with real $t_s$ (no phase), and with complex $t_x=|t_x| \exp(-i \phi_{\rm AB} m)$ dependent on $m$.  As shown in Fig.~\ref{Setup}D, the sum of the tunneling phases around any individual plaquette is $\phi_{\rm AB}$. 

We implemented a 2-D lattice by combining a conventional optical lattice to define the long axis of our system (${\bf e}_x$ direction, with sites labeled by $j$), with three sequentially coupled internal ``spin'' states  to define the short axis of our system (${\bf e}_s$ direction, with just three sites labeled by $m\in\left\{-1,0,+1\right\}$); in parallel to the work described here, an analogous scheme has been realized for fermionic ytterbium\cite{Mancini2015}.  As in Fig.~\ref{Setup}C, this system effectively has an infinite repulsive potential for $|m| \geq 2$, allowing for the formation of robust edge states.  In each band of our engineered lattice (Fig.~\ref{Setup}E), the momentum along $\ex$ specifies the position in $\es$ (denoted by color on the curves), just as for 2-D electrons in Landau levels.  

\paragraph*{Technique}

As shown in Fig.~\ref{Setup}A-B, we used $\Rb87$ BECs in the $f=1$ ground state hyperfine manifold\cite{Lin2009}, confined in an optical dipole potential from two $1064\nm$ laser beams aligned along $\ex$ and $\ey$,  with trap frequencies $(\omega_x,\,\omega_y,\,\omega_z)/2\pi=(50,40,110)\Hz$.  We adiabatically\footnote{We linearly increased the lattice potential from $0$ to $\sim6 E_L$ in $200\ms$.} loaded such BECs onto the sites of the 1-D optical lattice formed by a pair of $\lambda_L = {1064.46\nm}$ laser beams counter-propagating along ${\bf e}_x$.  This laser defines the single-photon recoil momentum $\hbar k_L = 2\pi\hbar/\lambda_L$ and recoil energy $E_L = \hbar^2 k_L^2/2m_{\rm Rb}$, where $m_{\rm Rb}$ is the atomic mass.  The lattice depth $V\approx 6 E_L$ gave a hopping strength $|t_x|\approx0.05E_L$ along ${\bf e}_x$.

We then coupled the three $\ket{m}$ sub-levels either with two-photon Raman transitions or an rf magnetic field.  The $\lambda_R = 790.04\nm$ Raman lasers also counter-propagated\footnote{The Raman lasers counter-propagated at an angle $\theta = 0.75^{\circ}$ away from $\ex$, effectively reducing $k_R$ by $\approx 1\times10^{-4}$ and introducing a quadratic Zeeman like term with strength $-0.001 E_L$.  Both of these effects are negligible in our experiment.}  along $\ex$, with wave-vector $k_R = \pm 2\pi/\lambda_R$; the rf coupling effectively had $k_R=0$.  Either field gave a laboratory tunable effective tunneling strength $|t_s|\sim |t_x|$ along the spin ``direction'' ${\bf e}_s$,  proportional to the Rabi frequency $\Omega_R$.  The $2\hbar k_R$ momentum imparted from these transitions resulted in a spatially periodic phase factor $\exp(i 2 k_R x) = \exp(i\phi_{\rm AB}j)$ accompanying the change in $m$, where $\phi_{\rm AB}/2\pi= k_R/k_L\approx\pm4/3$ for Raman coupling and $\phi_{\rm AB}/2\pi=0$ for rf coupling\cite{Boada2012,Celi2014}.  The sign of $\phi_{\rm AB}$ was controlled by the relative detuning of the Raman lasers (see Fig.~\ref{Setup}A-B). 

Figure~\ref{Setup}E shows the band structure for our system featuring the three Hofstadter bands expected for $\phi_{\rm AB}/2\pi\approx4/3$, with our boundary conditions.  The system is most easily understood after making the local gauge transformation $\ket{j,m} \rightarrow \exp\left(-i \phi_{\rm AB} j m\right)\ket{j,m}$  which transfers the Peierls phase from $t_s$ into $t_x$.  This fully maps our system to Eq.~(1), and as shown in Fig.~\ref{Setup}D the result is an effective flux $\Phi/\Phi_0 = \phi_{\rm AB}/2\pi$ per plaquette.

\paragraph*{Experiment} 

%
%

We began our experiments by directly imaging adiabatically loaded eigenstates of the ground Hofstadter band with either $\phi_{\rm AB}/2\pi =0$ or $\approx4/3$, in an isotropic lattice with $|t_x|\approx |t_s|$.  After preparation, we used a measurement procedure common to all experiments: we simultaneously removed all potentials and coupling fields ($t_{\rm off} < 1\us$), which returned the atoms to bare spin and momentum states.  The atomic cloud expanded for a $\approx18\ms$ time-of-flight (TOF) period.  During TOF a $2\ms$ magnetic gradient pulse was applied, Stern-Gerlach separating the three $\ket{m}$ states.  The resulting 2-D column density was recorded using standard absorption imaging techniques, giving the normalized momentum distributions $n_m(k_x)$ with perfect single lattice site resolution along $\es$.

\begin{figure}[h!]
  \centering
    \includegraphics{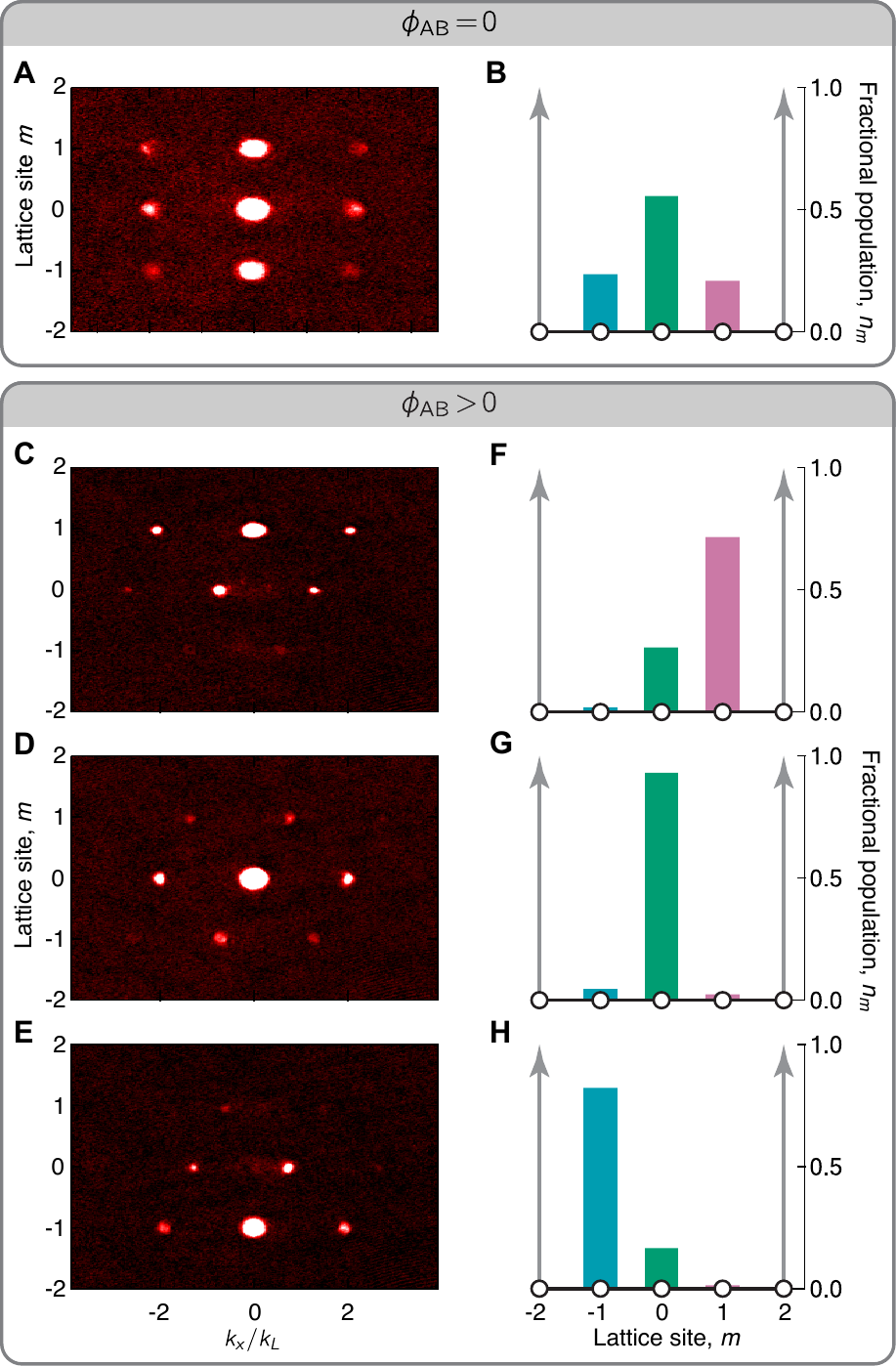}
    \caption{Adiabatically loaded eigenstates. Left panels: Site-resolved, normalized momentum distributions $n_m (k_x)$ obtained by absorption imaging.  Right panels: fractional population $n_m$.  (\textbf{A-B}) Atoms loaded into the single ground state present for $\phi_{\rm AB} = 0$ show the expected separable behavior along $\ex$ and $\es$.  (\textbf{C-H}) Atoms loaded into the upper edge, bulk, and lower edge states present for $\phi_{\rm AB}/2\pi\approx 4/3$ demonstrate the coupling between $\ex$ and $\es$ and localization along $\es$.}
    \label{EdgeStates}
\end{figure}

Figure~\ref{EdgeStates}A contains typical data with $\phi_{\rm AB}=0$, where we adiabatically loaded into the ground state and observed $n_m(k_x)$.  The fractional population $n_m=\int {\rm d} k_x n_m(k_x)$ resembles (Fig.~\ref{EdgeStates}B) that of a particle in a discretized box along $\es$, while the momentum distributions, typical for atoms in an optical lattice\cite{Greiner2001}, have the same profile for each $m$ site. This demonstrates that the two directions are uncoupled at $\phi_{\rm AB} = 0$.  

The data in Fig.~\ref{EdgeStates}C-E, with $\phi_{\rm AB} \neq 0$, are qualitatively different as a function of both $k_x$ and $m$.  These differences can be understood in analogy with a 2-D electron in a perpendicular magnetic field, confined in one dimension with hard walls. Along the confined direction the wavefunction is localized to the scale of the magnetic length $\ell_B = \sqrt{\hbar/qB}$, with center position at $k_x\ell_B^2$ in the bulk, where $\hbar k_x$ is the electron's canonical momentum.  For large $|k_x|$, the electron becomes localized near the edges, lifting the degeneracy of the otherwise macroscopically degenerate Landau levels.  Each of these points finds an analogue in our observations (Fig.~\ref{Setup}E, Fig.~\ref{EdgeStates}F-H).  In our system the magnetic length $\ell_B^* = \sqrt{3/2\pi}\approx0.7$ in units of lattice period is of order unity\footnote{Since all physics is $2\pi$ periodic in the acquired phase, our flux $\phi_{\rm AB}/2\pi\approx4/3$ is equivalent to $\phi_{\rm AB}\approx2\pi/3$.  For the function of estimating the magnetic length -- a continuum concept -- it is suitable to use $\ell_B^*\approx\sqrt{3/2 \pi}$.}, significantly narrowing the bulk state (Fig.~\ref{EdgeStates}G) as compared to the $\phi_{\rm AB} = 0$ case (Fig.~\ref{EdgeStates}B).  In addition, we observed the appearance of states localized at the system's edges (Fig.~\ref{EdgeStates}F,H), which are completely absent when $\phi_{\rm AB} = 0$.  These localized edge states are the analogue to the current carrying edge states in fermionic IQHE systems. 

%
%

\begin{figure*}[h!]
\centering
    \includegraphics{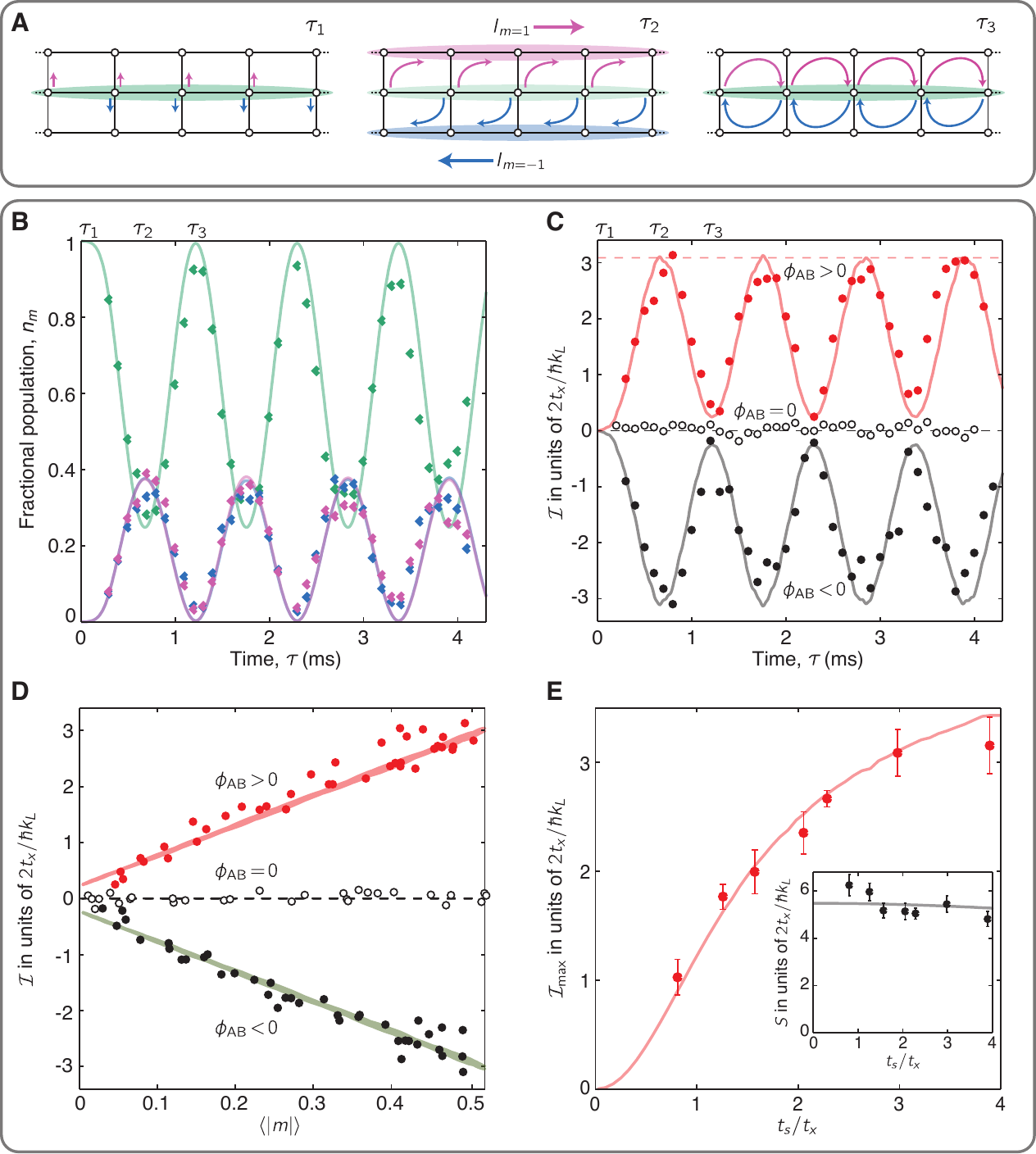}
    \caption{(\textbf{A}) Development of chiral edge currents. A system prepared at site $m=0$ at time $\tau_1=0$ obtains a chiral edge current at $\tau_2$ which returns to zero at $\tau_3$. (\textbf{B}) Fractional population versus time for atoms initialized at $m=0$ at time $\tau=0$.  The dynamics for $\pm\phi_{\rm AB}$ are the same.  (\textbf{C}) Chiral edge current $\mathcal{I}$ versus time.  Data shown in red, black, and empty circles are taken for positive, negative, and zero $\phi_{\rm AB}$, respectively. (\textbf{D}) $\mathcal{I}$ plotted against $ \langle |m|\rangle$. The solid curves (theory) use parameters $(\hbar\Omega_R,\,V,\,\delta,\,\epsilon)=(0.73, 6, 0.001, 0.05)E_L$ determined from B for $\phi_{\rm AB}\neq0$, giving $t_s=0.14E_L$, and $(\hbar\Omega_R, \, V, \, \delta, \, \epsilon)=(0.47, 6, -0.01, 0.05)E_L$ for $\phi_{\rm AB}=0$. (\textbf{E}) Maximum edge current versus asymmetry ($t_s/t_x$).  Inset: slope $S$ (taken from data as in D) is nearly independent of $t_s/t_x$.}
   \label{EdgeCurrent}
\end{figure*}

Having described the static properties of this system, we now turn to dynamics.  We loaded our systems on to different $m$ sites with $t_s=0$, and then abruptly\footnote{We commanded linear ramps to the Raman lasers giving a $300\us$ turn-on time.  This time scale was adiabatic with respect to the $\approx 5 E_L$ band spacing in the 1-D optical lattice, yet nearly instantaneous with respect to the $\approx 4 t_x$ magnetic band width.  Still, this non-zero turn-on time gave rise to the small offset in Fig.~\ref{EdgeCurrent}D, which is correctly predicted by our full theory.} turned on $t_s$, allowing tunneling along $\es$.  The resulting initial states all consisted of coherent superpositions of magnetic band eigenstates with crystal momentum $q_x/k_L=-m\phi_{\rm AB}/\pi$, which began to coherently tunnel along $\es$ and experienced an associated Lorentz force along $\ex$.  Atoms initialized on the bulk $m=0$ site demonstrated a dynamic Hall effect. Those starting on the edge sites became cold-atom analogues to edge magnetoplasmons: they began cyclotron orbits, were reflected from the hard wall, and skipped down one edge or the other.

The dynamics of atoms initialized in the bulk (on the $m=0$ site) are presented in Fig.~\ref{EdgeCurrent}.  As schematically illustrated in Fig.~\ref{EdgeCurrent}A and plotted in Fig.~\ref{EdgeCurrent}B, a balanced population oscillated in and out of the the originally empty $m=\pm1$ sites as a function of time $\tau$.  When $\phi_{\rm AB} \neq 0$, this motion drove transverse, i.e. Hall, edge currents $I_{m=\pm1}(\tau)=n_m(\tau) \langle V_m\rangle$ along $\ex$ (Fig.~\ref{EdgeCurrent}A), where $\langle V_m\rangle$ is the mean velocity of atoms on site $m$ along $\ex$.  

As shown in Fig.~\ref{EdgeCurrent}C, a chiral current $\mathcal{I}=I_1-I_{-1}$ developed with overall sign following that of $\phi_{\rm AB}$.  As atoms tunneled to the edges they acquired a transverse velocity controlled by two parameters: $\phi_{\rm AB}$ set the crystal momentum acquired while tunneling, and $t_x$ gave the natural unit of velocity $2 t_x / \hbar k_L$.  This led to the observed in-phase oscillation of $\mathcal{I}$ and the combined $m=\pm1$ populations $\langle |m| \rangle$.

This synchronous oscillation implies a linear dependance of $\mathcal{I}$ on  $\langle |m| \rangle$ whose slope we label $S$, plotted by the red and black symbols in Fig.~\ref{EdgeCurrent}D.  We confirmed the system's chirality by inverting $\phi_{\rm AB}$ and verifying that $S$ changed sign.  For comparison, we repeated the experiment with $\phi_{\rm AB}=0$, and observed no chiral current (empty symbols).  These data are all in good agreement with our theory (curves) using parameters obtained from fits to Fig.~\ref{EdgeCurrent}B.

The dependence of the chiral current on the tunneling anisotropy $t_s/t_x$ shown in Fig.~\ref{EdgeCurrent}E is reminiscent of the optical lattice experiments in Ref.~\cite{Atala2014}.  The chiral current remained linear in $\langle |m|\rangle$, with slope $S$ essentially constant\footnote{The slight downward curvature in our theoretical model, plotted by the grey curve, evidences 5\%-level corrections proportional to $|t_s|^2$ to the tight binding model [Eq.~(1)], owing to mixing with excited Bloch bands of the 1-D optical lattice.} in units of $2 t_x / \hbar k_L$; this demonstrates a new kind of dynamic Hall effect.    In contrast, the peak edge current $\mathcal{I}_{\rm max}$ (pink dashed line in Fig.~\ref{EdgeCurrent}C) strongly depends on $t_s/t_x$ (Fig.~\ref{EdgeCurrent}E), increasing from zero then reaching saturation.  For small $t_s/t_x$, few atoms tunneled giving a correspondingly small $\mathcal{I}_{\rm max}$, while as $t_s/t_x$ increased $\mathcal{I}_{\rm max}$ began to saturate as essentially all atoms tunnelled\cite{Hugel2014}. 

\begin{figure*}[]
  \centering
    \includegraphics{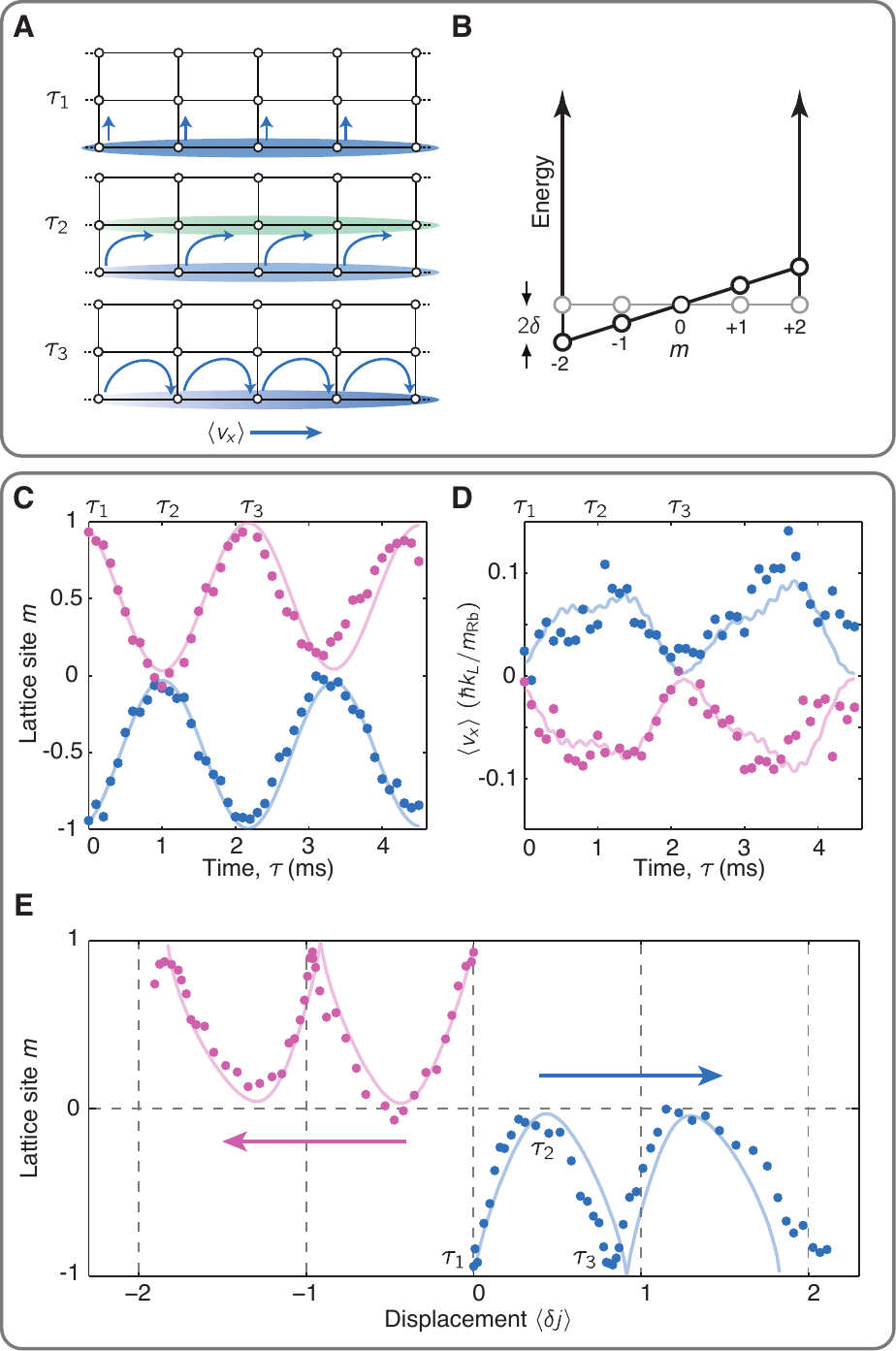}
    \caption{Skipping orbits.  (\textbf{A-B}) Schematic of dynamics starting from $m=-1$ including a nonzero detuning which tilted the lattice along $\es$. (\textbf{C}) Mean displacement $\langle m \rangle$ versus time $\tau$ for excitations on both edges. Data for systems initialized on $m=1$ ($m=-1$) are depicted by pink (blue) circles. (\textbf{D}) Group velocity along $\ex$ versus time for both edge excitations. Numerical simulations (solid curves) use parameters $(\hbar\Omega_R,\,V,\,\delta,\,\epsilon)=(0.58,5.2,\pm0.087,0.13)E_L$ from fits to population dynamics like C. In both cases, the potential gradient along $\es$ from $\delta$ was selected so the initial site ($m=\pm1$) had the lowest energy. (\textbf{E}) Edge magnetoplasmon trajectories, where the displacement $\langle\delta j(\tau)\rangle$ was obtained by integrating $\langle v_x(\tau)\rangle/a$.}
    \label{SkippingOrbit}
\end{figure*}

We then moved from studying bulk excitations to edge excitations by launching edge magnetoplasmons: superpositions of edge states across magnetic bands with crystal momentum $q_x/k_L=\mp\phi_{\rm AB}/\pi$.   We created them on either edge, with the potential tilted along $\es$ as in Fig.~\ref{SkippingOrbit}A-B, such that the initially occupied site was at the potential minimum.  In Fig.~\ref{SkippingOrbit}C-D, we plot the time evolving average position $\langle m(\tau)\rangle$ along $\es$ and velocity $\langle v_x(\tau)\rangle=\sum_m I_m$ along $\ex$.  Data shown in pink/blue solid circles are for initial sites $\langle m(\tau=0)\rangle=\pm1$, both of which evolved periodically in time but with opposite velocities.  The spatial trajectories are illustrated in Fig.~\ref{SkippingOrbit}E, where we obtained the displacement $\langle \delta j(\tau)\rangle$ by directly integrating $\langle v_x(\tau)\rangle/a$, where $a=\lambda_L/2$ is the lattice period. These data clearly show edge magnetoplasmons with their chiral longitudinal motion, and constitute the first experimental observation of their edge localization and transverse skipping motion. 

\paragraph*{Conclusion}

We realized a large artificial magnetic field --- with $\phi_{\rm AB} /2\pi\approx 4/3$ flux per plaquette --- in a 2-D lattice geometry.  The use of internal atomic states as one of the lattice directions made single site ``spatial'' resolution along that axis straightforward, enabling our direct observation of chiral edge states,  a dynamical quantum Hall effect, and edge magnetoplasmons.  This and related approaches\cite{Mancini2015} have the technical advantage over other techniques for creating artificial gauge fields in that minimal Raman laser coupling is required (typically 10 to 50 times less than previous experiments using Raman coupling\cite{Lin2009b}) thereby minimizing heating from spontaneous emission and enabling many-body experiments which require negligible heating rates.  Lifetimes from spontaneous emission with this technique are in excess of 10 seconds, while all other approaches for creating large artificial gauge fields have lifetimes well below 1 second\cite{Struck2012,Aidelsburger2013,Miyake2013,Jotzu2014}.  

With our hard-wall potential, a realization of the Laughlin charge pump\cite{Laughlin1981} is straightforward: as particles accelerate along $\ex$, mass moves from one edge to the other in the orthogonal direction $\es$.  Remarkably, extending our technique to periodic boundary conditions ---coupling together the $\ket{m=\pm1}$ states--- should produce systems exhibiting a fractal Hofstadter spectrum\cite{Celi2014}, even given a three-site extent along ${\bf e}_s$.  Going beyond conventional condensed matter realities, the flexibility afforded by directly laser-engineering the hopping enables the creation of M\"obius strip geometries: topological systems with just one edge\cite{Boada2014}.

\bibliography{SynDim}

\bibliographystyle{Science}


\begin{scilastnote}
\item[{\bf Acknowledgments}] This work was partially supported by the ARO's Atomtronics MURI, by the AFOSR's Quantum Matter MURI, NIST, and the NSF through the PFC at the JQI.  B.K.S. is a NIST-NRC Postdoctoral Research Associate. L.M.A. was supported by the NSF GRFP.
\item[{\bf Author contributions}] H.-I~L. and B.K.S contributed equally to this project.  All authors excepting I.B.S contributed to the data taking effort.  B.K.S. and L.M.A. configured the apparatus for this experiment.  H.-I~L. lead the team's effort on all aspects of the edge current and skipping orbit measurements.  H.-I~L., L.M.A., and B.K.S. analyzed data. B.K.S., H.-I~L., L.M.A, and I.B.S. performed numerical and analytical calculations.  All authors contributed to writing the manuscript.  I.B.S. proposed the initial experiment.
\end{scilastnote}


\clearpage

\end{document}